\documentclass[12pt]{article}
\usepackage{amsfonts,amsmath,amssymb}
\usepackage{graphicx}
\usepackage{cite}
\usepackage{bm}

\allowdisplaybreaks[1]

\def\prt{\partial}

\newcommand{\be}{\begin{equation}}
\newcommand{\ba}{\begin{eqnarray}}
\newcommand{\ea}{\end{eqnarray}}
\newcommand{\ee}{\end{equation}}

\newcommand{\GN}{G_4}
\newcommand{\R}{\mathbb{R}}
\newcommand{\Z}{\mathbb{Z}}

\begin{document}

\begin{titlepage}

\begin{flushright}
arXiv:0811.4177 [hep-th]\\
KUNS-2169\\
YITP-08-88
\end{flushright} 

\vspace{0.1cm}

\begin{center}
  {\LARGE 
  Holographic Duals of Kaluza-Klein Black Holes
  }
\end{center}
\vspace{0.1cm}
\begin{center}

         Tatsuo A{\sc zeyanagi}$^{\dagger}$\footnote
           {
E-mail address : aze@gauge.scphys.kyoto-u.ac.jp}, 
         Noriaki O{\sc gawa}$^{\S}$\footnote
         {
E-mail address : noriaki@yukawa.kyoto-u.ac.jp} and    
         Seiji T{\sc erashima}$^{\S}$\footnote
           {
E-mail address : terasima@yukawa.kyoto-u.ac.jp}

\vspace{0.3cm}

${}^{\dagger}$
{\it Department of Physics, Kyoto University,\\
Kyoto 606-8502, Japan}\\

${}^{\S}$
{\it Yukawa Institute for Theoretical Physics, Kyoto University,\\
     Kyoto 606-8502, Japan}\\

\end{center}

\vspace{1.5cm}

\begin{abstract}
We apply Brown-Henneaux's
method to the 5D extremal rotating Kaluza-Klein black holes
essentially following the calculation of the Kerr/CFT correspondence,
which is not based on supersymmetry nor string theory.
We find that
there are two completely different Virasoro algebras
that can be obtained as the asymptotic symmetry algebras
according to appropriate boundary conditions.
The microscopic entropies are
calculated by using the Cardy formula for both boundary conditions and
they perfectly agree with the Bekenstein-Hawking entropy.
The rotating Kaluza-Klein black holes contain a 4D dyonic
Reissner-Nordstr\"om black hole and Myers-Perry black hole.
Since the D-brane configurations
corresponding to these black holes are known, we expect that our analysis 
will shed some light on deeper understanding of chiral $CFT_2$'s
dual to extremal black holes.
\end{abstract}

\end{titlepage}

\tableofcontents

\section{Introduction}
\label{introduction}
More than ten years have passed since the Bekenstein-Hawking entropy 
was calculated microscopically by using string theory 
\cite{Strominger:1996sh}. 
The essential strategy for microscopic entropy counting is to 
construct a black hole geometry by D-branes and then  
to map the problem of the black hole 
to a dual boundary field theory on the D-branes.
Then by using the BPS properties, 
we can count the microscopic entropy of the black hole in 
the weak coupling field theory.

After this accomplishment, this strategy is 
applied to various kinds of black holes and even other 
black objects like black rings. In addition to this, 
the $AdS/CFT$ correspondence, a kind of duality between gravity and field theory, 
inherits this feature and refines or makes this strategy more 
stringent in certain setups 
\cite{Maldacena:1997re, Gubser:1998bc, Witten:1998qj}.  

In the context of $AdS/CFT$, black hole microstates 
are investigated mainly by using $AdS_3/CFT_2$ correspondence since 
$AdS_3$ appears as a near-horizon geometry of 5D extremal black 
holes and $CFT_2$ is well analyzed.
When we consider 4D extremal black holes, on the other hand, 
$AdS_2$ naturally appears as a near-horizon geometry. However, 
$AdS_2/CFT_1$ is still mysterious because we do not know 
$CFT_1$ well. For some context, $CFT_1$ is regarded as 
conformal quantum mechanics (CQM)
\cite{Michelson:1999zf,Michelson:1999dx,Maloney:1999dv,BrittoPacumio:2000sv,%
BrittoPacumio:2001kc,BrittoPacumio:1999ax,Gaiotto:2004ij},
while it is as $2$D chiral conformal field theory
\cite{Maldacena:1998bw, Strominger:1998yg,Spradlin:1999bn,Azeyanagi:2007bj}. 
In this paper we investigate the latter. 

Going back to the history of the relation between conformal field theory 
and gravity, in $1986$, Brown and Henneaux succeeded in identifying 
generators of Virasoro algebras living on the boundary of $AdS_3$ spacetimes
with the diffeomorphisms preserving a certain boundary condition
\cite{Brown:1986nw}. 
They found that the central charge of the Virasoro algebra
is related to the $AdS_3$ radius.  
This method is applied to microscopically derive the entropy of
the BTZ black hole \cite{Strominger:1997eq}.      
Subsequently, related methods were developed
\cite{Carlip:1998wz,Carlip:1999cy,Park:2001zn,Carlip:2002be},
and in this direction many works were done
(for example, \cite{Cadoni:1998sg,Cadoni:1999ja,Castro:2008ms,Hotta:2008xt}).

Recently, the entropy of the 4D extremal Kerr 
black hole was microscopically calculated by 
applying Brown-Henneaux's method to the near-horizon 
geometry of it \cite{Guica:2008mu}. 
This is interesting partly because
their calculation was not based on supersymmetry nor string theory.
However, the Brown-Henneaux's method does not tell us 
what is the boundary theory, thus
it is desired to find the boundary theory explicitly. 
Microscopic counting of neutral black holes like Kerr black hole, 
is very difficult in that we do not know 
how to realize this neutrally charged geometry by D-branes.  
In \cite{Horowitz:2007xq}, the entropy of the 4D extremal Kerr 
black hole was microscopically calculated by relating 
it to a rotating Kaluza-Klein (KK) black hole solution formulated  
in \cite{Gibbons:1985ac,Rasheed:1995zv,Matos:1996km,Larsen:1999pp}.
This solution itself is meaningful in that 
it includes various classes of black holes like 
the 4D dyonic Reissner-Nordstr\"om black hole. 
Since the rotating KK black holes are constructed 
as a D0-D6 bound state \cite{Itzhaki:1998ka},
their entropies are calculable microscopically 
from the corresponding brane configurations
\cite{Emparan:2006it,Emparan:2007en}. 

In this paper, we would like to apply Brown-Henneaux's 
method to the 5D extremal rotating KK black holes and 
calculate its entropy microscopically.
We essentially follow the calculation of \cite{Guica:2008mu}. 
First we take near-horizon limit for the extremal rotating KK black holes.
For this near-horizon geometry,
it is shown that
two completely different Virasoro algebras 
can be obtained as the asymptotic symmetry algebras,
according to appropriate boundary conditions.
These algebras will act on the Hilbert states of 
the holographic duals of the extremal rotating KK black hole.
Although these dual boundary theories are not specified explicitly,
the microscopic entropies are
calculated using the Cardy formula
for both boundary conditions and  
they perfectly agree with the Bekenstein-Hawking entropy.
Therefore we expect that the two boundary conditions 
correspond to two different consistent holographic duals.
Since the D-brane configurations
corresponding to these black holes are known,
we expect that our analysis will shed some light
on deeper understanding of chiral $CFT_2$'s dual to extremal black holes.

The organization of this paper is as follows. 
In section \S\ref{kerr}, 
we review the work \cite{Guica:2008mu}.
In section \S\ref{review_rkk}, 
we review the rotating KK black holes.
In section \S\ref{near_horizon_rkk} we take the 
near-horizon limit of the extremal rotating KK black holes.
In section \S\ref{brown_henneaux}, we impose boundary 
conditions in the asymptotic region and determine 
the diffeomorphisms which preserve the boundary conditions.
We construct Virasoro generators in two ways since 
the rotating KK black holes contain two $U(1)$ fibers.
Then we find the central charges of these Virasoro algebras
living on the boundary. In section \S\ref{temperature}, we 
relate the physical parameters of the rotating KK black hole 
to the effective temperatures of
the dual chiral $CFT_2$'s and then, in section \S\ref{entropy},
we apply the Cardy formula to calculate the entropy microscopically.
Finally, in \S\ref{coclusion_and_discussion}, we summarize 
our conclusion and give some discussion.

\section{A Brief Review of Kerr/CFT Correspondence} 
\label{kerr}

In \cite{Guica:2008mu}, it is proposed that the 
4D extremal Kerr black hole is dual to a 
chiral $CFT_2$. 
In the extremal limit \cite{Bardeen:1999px}, 
the near-horizon geometry is of the form 
\begin{align}
ds^2 &= 2\GN J\Omega^2
\Bigl(
-(1+r^2)d\tau^2+\frac{dr^2}{1+r^2}
+d\theta^2+\Lambda^2(d\varphi+rd\tau)^2
\Bigr), \label{nhek}\\ 
&\Omega^2=\frac{1+\cos^2\theta}{2},
\quad \Lambda = \frac{2\sin\theta}{1+\cos^2\theta},
\end{align}
where $J$ is the angular momentum and
$\GN$ is the 4D Newtonian constant.
This angular momentum is related to the ADM mass $M$
of the black hole as $J=\GN M^2$ in the extremal limit.
For a fixed $\theta$, this geometry is the same as 
a quotient of a warped $AdS_3$, which is analyzed 
in the context of topological massive gravity in 3D \cite{Anninos:2008fx}. 

A generator of the Virasoro algebra of the chiral $CFT_2$ is 
identified with a class of diffeomorphism 
which preserves an appropriate boundary condition on 
the near-horizon geometry.
Then it is found that a nontrivial part 
of the diffeomorphism, or the asymptotic symmetry group (ASG),
contains diffeomorphisms of the form 
\begin{align}
\zeta_{n}=-e^{-in\varphi}\partial_{\varphi}-inre^{-in\varphi}\partial_r
\quad (n=0,\pm1,\pm2,\cdots).  
\label{kerr_zeta}
\end{align}
We notice that $\zeta_{n}$ contains $\partial_{\varphi}$, 
not $\partial_{\tau}$.

The diffeomorphisms $\zeta_{n}$
generate a Virasoro algebra without a central charge 
\begin{align}
i[\zeta_m, \zeta_n] =(m-n)\zeta_{m+n}.
\label{virasoro_zero}
\end{align}
By following the covariant formalism of the ASG
\cite{Barnich:2001jy, Barnich:2007bf, Abbott:1981ff,Iyer:1994ys,%
Anderson:1996sc, Torre:1997cd, Barnich:1994db, Barnich:2000zw,%
Barnich:2003xg, Compere:2007az},
a conserved charge $Q_{\zeta}$ associated with an element $\zeta$ is defined by
\begin{align}
Q_{\zeta} =\frac{1}{8\pi G_d}
\int _{\partial\Sigma} k_{\zeta}[h,\bar{g}], 
\label{charge_zeta}
\end{align}
where $\partial\Sigma$ is a spatial surface at infinity and
\begin{align}
k_{\zeta}[h,\bar{g}]
&=\frac{1}{2}
\Bigl[
\zeta_{\nu}D_{\mu}h -\zeta_{\nu}D^{\sigma}h_{\mu\sigma}+
\zeta^{\sigma}D_{\nu}h_{\mu\sigma} \nonumber\\
&\qquad +\frac{1}{2}h D_{\nu}\zeta_{\mu}
-h_{\nu\sigma}D^{\sigma}\zeta_{\mu}
+\frac{1}{2}h_{\nu\sigma}(D_{\mu}\zeta^{\sigma}+D^{\sigma}\zeta_{\mu})
\Bigr]
*_d\!(dx^{\mu}\wedge dx^{\nu}).
\label{current_zeta}
\end{align}
Here $d$ is the spacetime dimension ($d=4$ in the current case),
$*_d$ represents the Hodge dual,
$\bar{g}_{\mu\nu}$ is the metric of the background geometry 
(\ref{nhek}) and $h_{\mu\nu}$
is deviation from it.
We also notice that the covariant derivative
in (\ref{current_zeta}) is defined by using $\bar{g}_{\mu\nu}$.
In addition to a charge $Q_{\zeta_n}$ associated with $\zeta_n$, 
there exists a charge $Q_{\partial_{\tau}}$ associated with 
$\partial_{\tau}$.
Since it measures deviation from the extremality, 
it is fixed to zero in this case.

Then let us consider the Dirac bracket of $Q_{\zeta_n}$
under the constraint $Q_{\partial_{\tau}}=0$.
It is determined by considering transformation property of the charge 
$Q_{\zeta_n}$ under a diffeomorphism generated by $\zeta_m$. 
It then follows that
\begin{align}
\{Q_{\zeta_{m}},Q_{\zeta_n}\}_{Dirac}=Q_{[\zeta_{m},\zeta_{n}]}
+\frac{1}{8\pi G_4}\int_{\partial \Sigma}
k_{\zeta_{m}}[\mathcal{L}_{\zeta_{n}}\bar{g},\bar{g}].
\label{dirac_algebra}
\end{align}
By redefining the charge as $\hbar L_n=Q_{\zeta_n}+3J\delta_{n,0}/2$ 
and replacing the Dirac bracket $\{.,.\}$ 
by a commutator $-\frac{i}{\hbar}[.,.]$,
we see that $L_n$ satisfies a Virasoro algebra 
\begin{align}
[L_m,L_n] =(m-n)L_{m+n}+\frac{c}{12}m(m^2-1)\delta_{m+n,0},
\label{virasoro_central}
\end{align}
with the central charge $c=12J/\hbar$. 

The temperature of the dual chiral $CFT_2$ is, on the other hand, 
determined by
identifying quantum numbers in the near horizon geometry 
with those in the original geometry.
This method gives the temperature
\begin{align}
T=\frac{1}{2\pi}.
\end{align}  
From this, entropy of the 4D Kerr black hole is microscopically 
calculated via the Cardy formula as
\begin{align}
S_{\mathit{micro}}=\frac{\pi^2}{3}cT = \frac{2\pi J}{\hbar}, 
\end{align}
which agrees with the Bekenstein-Hawking 
entropy calculated macroscopically.

We note that 
this dual theory is not like $CFT$ in the usual $AdS/CFT$ correspondence,
because $\phi$ is space-like coordinate and 
the Virasoro algebra does not contain the time translation generator.
Moreover, the isometry of the near horizon geometry does not 
contain $SL(2,R)$ of the Virasoro algebra.
We will see that 
the Virasoro algebras of the holographic duals of 
the rotating KK black holes also have this properties.

\section{The Rotating Kaluza-Klein Black Holes}
\label{review_rkk}

In this section we review the rotating Kaluza-Klein
black holes \cite{Gibbons:1985ac,Rasheed:1995zv,Matos:1996km,Larsen:1999pp}.
This is the 5D-uplifted solution of rotating black holes
with both electric and magnetic charges
in the 4D Einstein-Maxwell-dilaton theory.
This solution includes
the dyonic ($P=Q$) Reissner-Nordstr\"om black hole
in the 4D Einstein-Maxwell theory and the 5D Myers-Perry black hole,
as special cases.%
\footnote{
Precisely speaking, it corresponds to the Myers-Perry black hole
on an orbifolded space $\R^{1,4}/\Z_{N_6}$,
with $N_6$ an even integer.
Details of the transformations are given in
\cite{Emparan:2007en}.
}
In terms of string theory,
it can be interpreted as a rotating D0-D6 bound state.

We consider the 4D Einstein-Maxwell-dilaton action
\begin{align}
S=\frac{1}{16\pi\GN}\int d^4 x \sqrt{-g}
\Bigl[
\mathcal{R}-2g^{\mu\nu}\prt_\mu\Phi\prt_\nu\Phi
-\frac{1}{4}e^{-2\sqrt{3}\Phi}g^{\mu\alpha}g^{\nu\beta}F_{\mu\nu}F_{\alpha\beta}
\Bigr].
\label{4d_action}
\end{align}
This theory is obtained by a usual Kaluza-Klein reduction of
the 5D pure Einstein gravity theory,
which is easier to deal with in many cases.
In this paper, we will always work on the 5D theory.%
\footnote{
The central charges and the entropy
we will obtain do not depend on the Kaluza-Klein radius $R$.
Therefore they would also be valid for very small $R$
compared to the black hole radius,
where the masses of the higher Kaluza-Klein modes become very large
and the description by \eqref{4d_action} is expected to be exact.
However, for $\zeta^y_n$ \eqref{zetayn},
which relate massive KK modes,
this limit $R\to 0$ is a little subtle.
}

In terms of the 5D pure Einstein gravity,
the rotating KK solution is written as
\begin{align}
ds_{(5)}^2=
\frac{H_2}{H_1}(R\,d\hat{y}+\bm{A})^2
-\frac{H_3}{H_2}(d\hat{t}+\bm{B})^2
+H_1\Bigl(\frac{d\hat{r}^2}{\Delta}+d\theta^2
+\frac{\Delta}{H_3}\sin^2\theta\,d\phi^2\Bigr),
\label{5d_rkk_metric}
\end{align}
in which
\begin{align}
\label{rkk_functions_start}
H_1 &= \hat{r}^2 + \mu^2j^2\cos^2\theta + \hat{r}(p-2\mu)+\frac{1}{2}\frac{p}{p+q}(p-2\mu)(q-2\mu) \nonumber\displaybreak[0]\\
&\quad+\frac{1}{2}\frac{p}{p+q}\sqrt{(p^2-4\mu^2)(q^2-4\mu^2)}\,j\cos\theta,
\displaybreak[2]\\
H_2 &= \hat{r}^2 + \mu^2j^2\cos^2\theta + \hat{r}(q-2\mu)+\frac{1}{2}\frac{q}{p+q}(p-2\mu)(q-2\mu) \nonumber\displaybreak[0]\\
&\quad-\frac{1}{2}\frac{q}{p+q}\sqrt{(p^2-4\mu^2)(q^2-4\mu^2)}\,j\cos\theta,
\displaybreak[2]\\
H_3 &= \hat{r}^2 + \mu^2j^2\cos^2\theta -2\mu\hat{r},
\displaybreak[2]\\
\Delta &= \hat{r}^2 + \mu^2j^2 - 2\mu\hat{r},
\displaybreak[2]\\
\bm{A} &= -\Biggl[ \sqrt{\frac{q(q^2-4\mu^2)}{p+q}}
 \Bigl(\hat{r}+\frac{p-2\mu}{2}\Bigr)
- \frac{1}{2}\sqrt{\frac{q^3(p^2-4\mu^2)}{p+q}}j\cos\theta \Biggr] H_2^{-1}d\hat{t} \nonumber\displaybreak[1]\\
&\quad+ \Biggl[ -\sqrt{\frac{p(p^2-4\mu^2)}{p+q}}(H_2+\mu^2j^2\sin^2\theta)\cos\theta \nonumber\displaybreak[0]\\
&\qquad\;\;\, +\frac{1}{2}\sqrt{\frac{p(q^2-4\mu^2)}{p+q}}
\Bigl\{p\hat{r}-\mu(p-2\mu)+\frac{q(p^2-4\mu^2)}{p+q}\Bigr\}j\sin^2\theta
\Biggr] H_2^{-1}d\phi,
\displaybreak[2]\\
\bm{B} &=\frac{1}{2}\sqrt{pq}\,\frac{(pq+4\mu^2)\hat{r}-\mu(p-2\mu)(q-2\mu)}{p+q}
H_3^{-1}j\sin^2\theta\,d\phi,
\label{rkk_functions_end}
\end{align}
where $\hat{y}\sim \hat{y}+2\pi$ and $R$ is the radius of
the Kaluza-Klein circle at $\hat{r}\to\infty$.%
\footnote{
We follow this form of the solution from  \cite{Larsen:1999pp}
and checked that this indeed satisfies 5D Ricci flat condition.
Note that there are some typos in \cite{Larsen:1999pp}.
}
After the Kaluza-Klein reduction along $\hat{y}$ direction,
we obtain a 4D black hole of the form
\begin{align}
ds_{(4)}^2 &=
-\frac{H_3}{\sqrt{H_1H_2}}(d\hat{t}+\bm{B})^2
+\sqrt{H_1H_2}\Bigl(\frac{d\hat{r}^2}{\Delta}+d\theta^2+\frac{\Delta}{H_3}\sin^2\theta\,d\phi^2\Bigr),
\label{4d_metric} \\
e^{2\Phi} &= R^2\sqrt{\frac{H_1}{H_2}},
\label{4d_dilaton} \\
\bm{A}_{(4)} &= \frac{1}{R}\bm{A}.
\label{4d_gaugefield}
\end{align}

The rotating KK solution has four parameters $(\mu,j,q,p)$,
which correspond to
four physical parameters of the reduced 4D black hole,
that is, the ADM mass $M$, angular momentum $J$,
electric charge $Q$ and magnetic charge $P$.
The explicit relations between these parameters are \cite{Larsen:1999pp}:
\begin{align}
\label{physical_parameters_start}
M &= \frac{p+q}{4\GN},\\
J &= \frac{\sqrt{pq}(pq+4\mu^2)}{4\GN(p+q)}\,j,\\
Q &= \frac{1}{2}\sqrt{\frac{q(q^2-4\mu^2)}{p+q}}, \\
P &= \frac{1}{2}\sqrt{\frac{p(p^2-4\mu^2)}{p+q}}.
\label{physical_parameters_end}
\end{align}
Here we set $J,Q,P \ge 0$ for simplicity.
The possible range of the parameters
for regular solutions are
\begin{align}
0 \le 2\mu \le q,p,
\quad 0 \le j \le 1,
\end{align}
and the black hole is extremal when we take $\mu \rightarrow 0$
with $j$ fixed finite.\footnote{
We can take another extremal limit
$j \rightarrow 1$ with $\mu$ fixed finite,
which corresponds to the so-called ``fast rotation'' case.
We will not consider this in this paper because the near horizon limit
would be difficult to analyze.
}
The outer/inner horizons are given by
\begin{align}
r_\pm = \mu\bigl(1\pm\sqrt{1-j^2}\bigr),
\label{horizon}
\end{align}
which lead to the Bekenstein-Hawking entropy
\begin{align}
S_{\mathit{BH}}=
\frac{\pi\sqrt{pq}}{2\GN\hbar}
\biggl(
\frac{pq+4\mu^2}{p+q}\sqrt{1-j^2}\, + 2\mu
\biggr).
\label{Sbh}
\end{align}
The Hawking temperature is
\begin{align}
\beta_H = \frac{1}{T_H} =
\frac{\pi\sqrt{pq}}{\mu\hbar}\biggl(
\frac{pq+4\mu^2}{p+q}+\frac{2\mu}{\sqrt{1-j^2}}
\biggr).
\label{betaH}
\end{align}
On the event horizon,
the rotational velocity $\Omega_{\phi}$, the electric potential $\Phi_E$
and the magnetic potential $\Phi_M$
in the 4D theory are
\begin{align}
\Omega_{\phi} &=
\frac{p+q}{\sqrt{pq}}\frac{2\mu j}{2\mu(p+q)+(pq+4\mu^2)\sqrt{1-j^2}}, 
\\
\Phi_E &= \frac{\pi T_H}{2\mu\GN\hbar}\sqrt{\frac{p(q^2-\mu^2)}{p+q}}
\Bigl(p+\frac{2\mu}{\sqrt{1-j^2}}\Bigr), \\
\Phi_M &= \frac{\pi T_H}{2\mu\GN\hbar}\sqrt{\frac{q(p^2-\mu^2)}{p+q}}
\Bigl(q+\frac{2\mu}{\sqrt{1-j^2}}\Bigr), 
\end{align}   
respectively. These physical quantities satisfy the first 
law of black hole thermodynamics:
\begin{align}
dM = T_H dS + \Phi_E dQ +\Phi_M dP - \Omega_{\phi} dJ. 
\end{align} 
We also notice that,
in terms of the 5D geometry,
the potential $\Omega_{y}$ corresponding to the 
Kaluza-Klein momentum is  
\begin{align}
\Omega_{y} = \frac{2\GN}{R}\,\Phi_E,
\label{kk_potential}
\end{align}
at the horizon.

Since we will focus on the extremal case in this paper,
we show the explicit form of 
\eqref{rkk_functions_start}-\eqref{rkk_functions_end},
\eqref{physical_parameters_start}-\eqref{physical_parameters_end}
and \eqref{Sbh} in that case here:
\begin{align}
H_1 &= \hat{r}^2 + p\hat{r} + \frac{1}{2}\frac{p^2q}{p+q}(1+j\cos\theta), \\
H_2 &= \hat{r}^2 + q\hat{r} + \frac{1}{2}\frac{pq^2}{p+q}(1-j\cos\theta), \\
H_3 &= \Delta = \hat{r}^2,\\
\bm{A} &= - \frac{q^{3/2}}{\sqrt{p+q}}
\Bigl[ \hat{r} + \frac{p}{2}(1-j\cos\theta) \Bigr] H_2^{-1}d\hat{t} \nonumber\\
&\quad+ \Biggl[ -\frac{p^{3/2}}{\sqrt{p+q}}\cos\theta
+\frac{1}{2}\sqrt{\frac{pq^2}{p+q}}
\Bigl(p\hat{r}+\frac{p^2q}{p+q}\Bigr)H_2^{-1}j\sin^2\theta
\Biggr] d\phi, \\
\bm{B} &= \frac{1}{2}\frac{(pq)^{3/2}}{p+q}\frac{j\sin^2\theta}{\hat{r}}
\,d\phi,\\
M &= \frac{p+q}{4\GN},\\
J &= \frac{(pq)^{3/2}}{4\GN(p+q)}\,j,\\
Q &= \frac{1}{2}\sqrt{\frac{q^3}{p+q}}, \\
P &= \frac{1}{2}\sqrt{\frac{p^3}{p+q}}, \\
\label{Sbh_extremal}
S_{\mathit{BH}}&=
\frac{\pi}{2\GN\hbar}\frac{(pq)^{3/2}}{p+q}\sqrt{1-j^2}
=\frac{2\pi}{\hbar}\sqrt{\frac{P^2Q^2}{\GN^2}-J^2}.
\end{align}

Before closing this section, we rewrite the entropy calculated above
in terms of the integer charges.
As discussed in \cite{Emparan:2007en}, 
the electric and the magnetic charges are quantized and written as 
\begin{align}
Q= \frac{2\GN\hbar N_0}{R}, \quad P = \frac{R N_6}{4},
\end{align}
where $N_6$ and $N_0$ are integer numbers which 
corresponds to the number of D6-branes and D0-branes, respectively,
if we embed the 5D KK black hole in IIA string theory with $R=g_s l_s$.
In addition to this, $J$ is also quantized as a result of 
the usual quantization of the angular momentum, so
\begin{align}
J = \frac{\hbar N_J}{2},
\end{align}
where $N_J$ is an integer.
By using these 
quantized quantities,
the entropy  \eqref{Sbh_extremal} in the extremal case is also written
as a quantized form: 
\begin{align}
S_{BH}= \pi\sqrt{N_0^2N_6^2-N_J^2}.
\end{align}

\section{Near-Horizon Geometry of Extremal Rotating\\ Kaluza-Klein Black Holes}
\label{near_horizon_rkk}

Here we derive the near horizon geometry of the extremal
($\mu=0$) rotating KK black holes.
It is already investigated in \cite{Astefanesei:2006dd,Kunduri:2008rs},
and related discussions about symmetries of
near-horizon geometries are given in \cite{Kunduri:2007vf}.

We first introduce near horizon coordinates as  
\begin{align}
t=\lambda\hat{t},
\quad r=\frac{\hat{r}}{\lambda},
\quad y=\hat{y}-\frac{1}{R}\sqrt{\frac{p+q}{q}}\,\hat{t},
\label{nearhorizon_coordinate}
\end{align}
while $\theta$ and $\phi$ are unchanged
although the black holes are rotating along the $\phi$ direction.
We will see that these coordinates are appropriate
for obtaining the near-horizon geometry.

The near-horizon limit is defined as $\lambda\to 0$
in \eqref{nearhorizon_coordinate}.
In this limit, under the extremal condition $\mu=0$,
the metric (\ref{5d_rkk_metric}) turns to
\begin{align}
ds^2&=
\frac{q}{p}\frac{1-j\cos\theta}{1+j\cos\theta}
 \Bigl(R dy+  \frac{2r}{q(1-j\cos\theta)}\sqrt{\frac{p+q}{q}}dt
 + \sqrt{\frac{p^3}{p+q}}\frac{j-\cos\theta}{1-j\cos\theta}d\phi \Bigr)^2 \nonumber\\
&\quad -\frac{2(p+q)}{q^2p(1-j\cos\theta)}
 \Bigl(r\,dt+\frac{(pq)^{3/2}}{2(p+q)}j\sin^2\theta\,d\phi \Bigr)^2 \nonumber\\
&\quad +\frac{p^2q(1+j\cos\theta)}{2(p+q)}
 \Bigl(\frac{dr^2}{r^2}+d\theta^2+\sin^2\theta\,d\phi^2\Bigr),
\label{metric_nh_poincare}
\end{align}
which we call {\it near-horizon extremal rotating Kaluza-Klein}
black hole (NHERKK) geometry.
This geometry  is a so-called squashed $AdS_2\times S^2$
with Kaluza-Klein $U(1)$ fibration on it.

We can  also rewrite \eqref{metric_nh_poincare}
by introducing \begin{align} 
\rho = \frac{r}{2PQ}, \;\;\;
z = \frac{R}{2P} y
\end{align}
as
\begin{align}
ds^2&=
2P^{4/3}Q^{2/3}
\Biggl[
\frac{2(1-j\cos\theta)}{1+j\cos\theta}
 \Bigl(dz+\frac{\rho}{1-j\cos\theta}dt
  + \frac{j-\cos\theta}{1-j\cos\theta}\,d\phi\Bigr)^2 \nonumber\\
&\quad -\frac{1}{1-j\cos\theta}\bigl(\rho\,dt + j\sin^2\theta\,d\phi\bigr)^2
 +(1+j\cos\theta)\Bigl(\frac{d\rho^2}{\rho^2}+d\theta^2+\sin^2\theta\,d\phi^2
\Bigr)\Biggr],
\label{rewritten_nherkk}
\end{align}
where
\begin{align}
z\sim z + 2\pi \frac{R}{2P}.
\end{align}
This metric is invariant under a transformation
$t \rightarrow C t , \rho \rightarrow \rho/C$ 
for an arbitrary constant $C$.

Both of these coordinates here are of Poincar\'e-type,
and they do not cover the whole space in a single patch.
Like the usual $AdS_2$ space, the whole NHERKK space is expected
to have two disconnected boundaries.
The boundary which is found in our coordinates is $r$ (or $\rho$) $\to\infty$,
which should be (a part of) one of the two.
Therefore when we would like to focus on one of the dual chiral $CFT_2$'s,
which lives on one of the two boundaries,
we can expect that our coordinates work well.

\section{Boundary Conditions and Central Charges}
\label{brown_henneaux}

In order to calculate the entropy of the rotating KK black holes,
we have to determine the central charges of the Virasoro algebras which
will act on the Hilbert spaces of the boundary theories.

\subsection{Boundary Conditions and Asymptotic Symmetry Groups}
By following the work by Brown and Henneaux \cite{Brown:1986nw},
the first step is to find some boundary condition
on the asymptotic variations of the metric
and the ASG which preserves this boundary condition nontrivially.
In fact, for the NHERKK metric \eqref{metric_nh_poincare},
we can see that (at least) two different boundary conditions are allowed
in order that some nontrivial ASG's exist.

\subsubsection{two boundary conditions for the metric}

Let us suppose that the metric is perturbed as 
$g_{\mu\nu}=\bar{g}_{\mu\nu}+h_{\mu\nu}$ where $\bar{g}_{\mu\nu}$ 
is \eqref{metric_nh_poincare} and $h_{\mu\nu}$ is some deviation from it.
One of the possible boundary conditions
for $h_{\mu\nu}$ is,
\begin{align}
\left(
\begin{array}{ccccc}
h_{tt}=\mathcal{O}(r^2)
&h_{tr}=\mathcal{O}(\frac{1}{r^2})
&h_{t\theta}=\mathcal{O}(\frac{1}{r})
&h_{t\phi}=\mathcal{O}(r)
&h_{ty}=\mathcal{O}(1)\\
h_{rt}=h_{tr}
&h_{rr}=\mathcal{O}(\frac{1}{r^3})
&h_{r\theta}=\mathcal{O}(\frac{1}{r^2})
&h_{r\phi}=\mathcal{O}(\frac{1}{r})
&h_{ry}=\mathcal{O}(\frac{1}{r}) \\
h_{\theta t}=h_{t\theta}
&h_{\theta r}=h_{r\theta}
&h_{\theta\theta}=\mathcal{O}(\frac{1}{r})
&h_{\theta\phi}=\mathcal{O}(\frac{1}{r})
&h_{\theta y} = \mathcal{O}(\frac{1}{r}) \\
h_{\phi t}=h_{t\phi}
&h_{\phi r} =h_{r\phi}
&h_{\phi\theta} =h_{\theta\phi}
&h_{\phi\phi}=\mathcal{O}(\frac{1}{r})
&h_{\phi y}=\mathcal{O}(1)\\
h_{yt}=h_{ty}
&h_{yr}= h_{ry}
&h_{y\theta}=h_{\theta y}
&h_{y\phi} =h_{\phi y}
&h_{yy}=\mathcal{O}(1)
\end{array}
\right),
\label{ry_boundary_condition}
\end{align}
and a general diffeomorphism which preserves 
(\ref{ry_boundary_condition}) is written as 
\begin{align}
\zeta &=
\Bigl[C_1+\mathcal{O}\bigl(\frac{1}{r^3}\bigr)\Bigr]\partial_t
+[-r\gamma'(y)+\mathcal{O}(1)]\partial_{r}
+\mathcal{O}\bigl(\frac{1}{r}\bigr)\partial_{\theta} \nonumber\\
&\quad +\Bigl[C_2 +\mathcal{O}\bigl(\frac{1}{r^2}\bigr)\Bigr]\partial_{\phi}
+\Bigl[\gamma(y)+\mathcal{O}\bigl(\frac{1}{r^2}\bigr)\Bigr]\partial_y,
\label{ry_general_diffeos}
\end{align}
where $C_1$, $C_2$ are arbitrary constants and
$\gamma(y)$ is an arbitrary function of $y$.
From this, 
the asymptotic symmetry group is generated
by the diffeomorphisms of the form%
\footnote{
\eqref{ry_general_diffeos} also includes $\zeta^t=\partial_t$,
but it is excluded from the ASG by requiring a constraint
which we will explain in \S\ref{Dirac_constraint},
}
\begin{align}
\zeta^\phi&=\partial_{\phi}, \label{phi_killing}\\
\zeta^y_\gamma&=\gamma(y)\partial_y-r\gamma'(y)\partial_r. \label{ry_generator}
\end{align}
Especially, (\ref{ry_generator})
generates the conformal group of the Kaluza-Klein circle. 
To see that it really obeys the Virasoro algebra,
we expand $\gamma(y)$ in modes and define 
$\gamma_{n}=-e^{-iny}$. Then we can see that 
$\zeta^y_{n}$, which are defined as
\begin{align}
\label{zetayn}
\zeta^y_{n}=\gamma_{n}\partial_{y}-
r\gamma_n'\partial_r,
\end{align} 
which obey the Virasoro algebra under the Lie bracket as 
\begin{align}
[\zeta^y_m,{\ }\zeta^y_n]_{Lie} =-i(m-n)\zeta^y_{m+n}.
\end{align}
We notice that 
the Virasoro generators are constructed from $r$ and $y$.
In other words, we see that the generators of the Virasoro algebra
act on only $y$-direction in the dual boundary field theory.
Thus it is very different from 
the usual holographic dual $CFT_2$ where the time direction $t$
play some role. It seems that we cannot describe
dynamical processes by using this Virasoro algebra,
but at least to calculate the entropy,
we can use the Virasoro algebra on the $y$-direction.

The other allowed boundary condition is,
\begin{align}
\left(
\begin{array}{ccccc}
h_{tt}=\mathcal{O}(r^2)
&h_{tr}=\mathcal{O}(\frac{1}{r^2})
&h_{t\theta}=\mathcal{O}(\frac{1}{r})
&h_{t\phi}=\mathcal{O}(1)
&h_{ty}=\mathcal{O}(r)\\
h_{rt}=h_{tr}
&h_{rr}=\mathcal{O}(\frac{1}{r^3})
&h_{r\theta}=\mathcal{O}(\frac{1}{r^2})
&h_{r\phi}=\mathcal{O}(\frac{1}{r})
&h_{ry}=\mathcal{O}(\frac{1}{r}) \\
h_{\theta t}=h_{t\theta}
&h_{\theta r}=h_{r\theta}
&h_{\theta\theta}=\mathcal{O}(\frac{1}{r})
&h_{\theta\phi}=\mathcal{O}(\frac{1}{r})
&h_{\theta y} = \mathcal{O}(\frac{1}{r}) \\
h_{\phi t}=h_{t\phi}
&h_{\phi r} =h_{r\phi}
&h_{\phi\theta} =h_{\theta\phi}
&h_{\phi\phi}=\mathcal{O}(1)
&h_{\phi y}=\mathcal{O}(1)\\
h_{yt}=h_{ty}
&h_{yr}= h_{ry}
&h_{y\theta}=h_{\theta y}
&h_{y\phi} =h_{\phi y}
&h_{yy}=\mathcal{O}(\frac{1}{r})
\end{array}
\right),
\label{rphi_boundary_condition}
\end{align}
and general diffeomorphism preserving \eqref{rphi_boundary_condition} can be
written as
\begin{align}
\zeta &=
\Bigl[C_1+\mathcal{O}\bigl(\frac{1}{r^3}\bigr)\Bigr]\partial_t
+[-r\epsilon'(\phi)+\mathcal{O}(1)]\partial_{r}
+\mathcal{O}\bigl(\frac{1}{r}\bigr)\partial_{\theta} \nonumber\\
&\quad +\Bigl[\epsilon(\phi)+\mathcal{O}\bigl(\frac{1}{r^2}\bigr)\Bigr]\partial_{\phi}
+\Bigl[C_3+\mathcal{O}\bigl(\frac{1}{r^2}\bigr)\Bigr]\partial_y,
\end{align}
where $C_1$, $C_3$ are arbitrary constants and
$\epsilon(\phi)$ is an arbitrary function of $\phi$.
The ASG is generated by
\begin{align}
\zeta^\phi_\epsilon&=\epsilon(\phi)\partial_\phi-r\epsilon'(\phi)\partial_r,
\label{rphi_generator}\\
\zeta^y&=\partial_y \label{y_killing}.
\end{align}
In exactly the similar manner as above,
we define $\epsilon_n =-e^{-in\phi}$ and
\begin{align}
\zeta^{\phi}_n = \epsilon_{n}\partial_{\phi}-
r\epsilon_n'\partial_r,
\end{align}
which obey the Virasoro algebra 
\begin{align}
[\zeta^{\phi}_m,{\ }\zeta^{\phi}_n]_{Lie} =-i(m-n)\zeta^{\phi}_{m+n}.
\end{align}
In this case
the Virasoro generator is constructed from $r$ and $\phi$.

Here we assume that these two boundary conditions 
correspond to two different realizations
of these classical theories in a full quantum theory of gravity, 
like string theory.
We will see that these boundary conditions indeed
lead to the correct black hole entropy.
This suggests that this very interesting phenomenon,
i.e. two completely different microscopic theories
for one geometry, may be true.

Note that in each of these boundary conditions,
the Virasoro symmetry comes from an enhancement
of one of the two $U(1)$ symmetries of the geometry,
with the other remaining unenhanced.
One may wonder whether
both of the $U(1)$ symmetries could be enhanced at the same time
with $\{\zeta^y_n\}$ and $\{\zeta^\phi_n\}$ living together in the ASG,
leading to a dual $CFT$ with two Virasoro symmetries.
For example, we can consider a
more relaxed boundary condition with
$h_{t\phi}=\mathcal{O}(r)$, $h_{ty}=\mathcal{O}(r)$,
$h_{\phi\phi}=\mathcal{O}(1)$, $h_{yy}=\mathcal{O}(1)$
and the other elements same as
\eqref{ry_boundary_condition} and \eqref{rphi_boundary_condition}.
This boundary condition indeed admits both $\{\zeta^y_n\}$ and $\{\zeta^\phi_n\}$.
However in this case, the ASG includes a wider class of diffeomorphisms,
some of which are not commutative with $\zeta^t$.
It means that we cannot fix the energy of the black hole,
therefore this boundary condition cannot be regarded as consistent.%
\footnote{
We thank Andrew Strominger,
for suggesting our mistakes at this point in the original version
of this paper.
}

\subsubsection{the energy constraint}
\label{Dirac_constraint}

Using (\ref{charge_zeta}),
the asymptotic
conserved charges corresponding to $\zeta^t$,
(\ref{ry_generator}) and (\ref{rphi_generator})
are defined by 
\begin{align}
Q_{\zeta^t}=\frac{1}{8\pi G_5}\int_{\partial \Sigma} k_{\zeta^t},
\quad Q_{\zeta^y_\gamma}=\frac{1}{8\pi G_5}\int_{\partial\Sigma} 
k_{\zeta^y_\gamma},
\quad Q_{\zeta^\phi_\epsilon}=\frac{1}{8\pi G_5}\int_{\partial\Sigma} 
k_{\zeta^\phi_\epsilon},
\end{align}
where $G_5=2\pi R\GN$ is the 5D Newtonian constant 
and $k_{\zeta}$ is defined by (\ref{current_zeta}).
Obviously (\ref{phi_killing}) and (\ref{y_killing}) are the cases of
$\epsilon=1$ in (\ref{rphi_generator}) and
$\gamma=1$ in (\ref{ry_generator}) respectively,
whose charges represent the variances of 
the angular momentum and the KK momentum.
Similarly to the case of the 4D Kerr black hole, 
the generator $\zeta^t$ is also included in the asymptotic symmetry algebra.
Therefore we set $Q_{\zeta^t}=0$ identically
in order that the black holes remain extremal.

\subsection{Central Charges}

Next we have to determine the centrally extended expressions of
the Virasoro algebras.
For the $(r,y)$-diffeomorphism (\ref{ry_generator}),
the second term on the right hand side of 
(\ref{dirac_algebra}) is calculated as
\begin{align}
\frac{1}{8\pi G_5}\int_{\partial \Sigma}
k_{\zeta^y_{m}}[\mathcal{L}_{\zeta^y_{n}}\bar{g},\bar{g}]
=
-i\frac{2}{\GN R}Q\bigl(P^2m^3+\frac{R^2}{2}m\bigr)\delta_{m+n,0}.
\end{align}
From this,
by defining the Virasoro operators $L^{y}_m$ of $y$-direction as
\begin{align}
\hbar L^{y}_{m}=Q_{\zeta^y_{m}}
+\frac{1}{\GN R}Q\Bigl(P^2+\frac{R^2}{2}\Bigr)\,\delta_{m,0},
\label{ry_virasoro_shift}
\end{align}
replacing $\{\cdot,{\ }\cdot\}_{Dirac}\to 
\frac{1}{i\hbar}[\cdot,{\ }\cdot]$ and substituting into
(\ref{dirac_algebra}), we finally have 
the Virasoro algebra 
\begin{align}
{}[L^{y}_{m},{\ } L^{y}_{n}] = (m-n)L^{y}_{m+n}
+\frac{2}{\GN R}QP^2(m^3-m)\,\delta_{m+n,0},
\end{align}
Therefore the central charge $c^y$ is
\begin{align}
c^{y} =
\frac{24}{\GN\hbar R}QP^2 = 3N_0N_6^2.
\label{ry_central_charge}
\end{align}

For the $(r,\phi)$-diffeomorphism (\ref{rphi_generator}),
we can calculate the central charge in a similar manner.
The second term on the right hand side of 
(\ref{dirac_algebra}) is calculated as
\begin{align}
\frac{1}{8\pi G_5}\int_{\partial \Sigma}
k_{\zeta^\phi_{m}}[\mathcal{L}_{\zeta^\phi_{n}}\bar{g},\bar{g}]
= -iJ(m^3-2m)\delta_{m+n,0}.
\end{align}
Then by defining $L^{\phi}_m$ as 
\begin{align}
\hbar L^{\phi}_{m}=Q_{\zeta^\phi_{m}}
-\frac{J}{2}\,\delta_{m,0},
\label{rphi_virasoro_shift}
\end{align}
we can see that the Virasoro algebra is
\begin{align}
[L^{\phi}_{m},{\ } L^{\phi}_{n}] =(m-n)L^{\phi}_{m+n}
+J\,(m^3-m)\delta_{m+n,0}.
\end{align}
Therefore the central charge $c^{\phi}$ is
\begin{align}
c^{\phi}
= \frac{12}{\hbar}J
= 6N_J.
\label{rphi_central_charge}
\end{align}

Before closing this section, we note that
the calculated central charges are integer numbers.
In particular for $(r,y)$ case, it is written by
$N_0$ and $N_6$ only.
This suggests that there is an underlying microscopic theory
which is obtained from some weak coupling theory on D-branes.
Since the central charge is completely different from
the one obtained in \cite{Emparan:2006it, Emparan:2007en},
it is highly interesting to investigate
the corresponding D-brane system.

\section{Temperatures}
\label{temperature}

In the previous section, we derived the central charge
of the Virasoro algebra for two cases. 
Next we have to determine the ``temperature'' $T$ of the 
chiral $CFT_{2}$ following \cite{Guica:2008mu},
since entropy is microscopically calculated by
the thermal representation of the Cardy formula
\begin{align}
S=\frac{\pi^2}{3}cT.
\label{cardy}
\end{align}

For this purpose, let us consider a free scalar field $\Psi$ 
propagating on (\ref{5d_rkk_metric}). It can be expanded as 
\begin{align}
\Psi=\displaystyle{\sum_{\omega,k,m,l}
e^{-i\omega \hat{t}+ik\hat{y}+im\hat{\phi}}}f_{m,l}(r,\theta),
\end{align}
where $\omega$ is the asymptotic energy of the scalar field, while 
$k$ and $m$ are Kaluza-Klein momentum in the $y$-direction and 
the quantum number corresponding to the angular velocity respectively. 
We also notice that $m,{\ }l$ label the spherical harmonics.
Using the coordinates of the near-horizon geometry 
(\ref{nearhorizon_coordinate}), we see that 
\begin{align}
e^{-i\omega\hat{t}+ik\hat{y}+im\hat{\phi}}
=e^{-in^{t}t+in^{y} y +in^{\phi} \phi},\\
n^t=\frac{1}{\lambda}\Bigl(
\omega - \frac{k}{R}\sqrt{\frac{p+q}{q}}
\Bigr), 
\quad n^{y}=k, \quad n^{\phi} = m.
\label{nlnr}
\end{align}
After tracing out the states inside the horizon,
we find that the vacuum state is expected to include 
a Boltzmann factor of the form
\begin{align}
e^{-\frac{\hbar(\omega-k\Omega_{y}+m\Omega_{\phi})}{T_{H}}}
=e^{-\frac{n^{t}}{T^{t}}-\frac{n^{y}}{T^{y}}-\frac{n^{\phi}}{T^{\phi}}}.
\label{boltzmann}
\end{align}
The temperatures $T^{t}$, $T^{y}$ and $T^{\phi}$ are
calculated as 
\begin{align}
T^t=\frac{T_H}{\hbar\lambda},
\quad T^y= 
\frac{T_H}{\hbar\bigl(\frac{1}{R}\sqrt{\frac{p+q}{q}}-
\Omega_{y}\bigr)}, \quad 
T^{\phi} = \frac{T_H}{\hbar\Omega_\phi},
\end{align}
where we used (\ref{nlnr}).
When $\mu,\lambda\to 0$ as ${\mu/\lambda}\to 0$,
we see that $T^t\to0$, while $T^y$ and $T^{\phi}$ are 
\begin{align}
T^y &=
\frac{\GN R}{4\pi P^2Q}\sqrt{\frac{P^2Q^2}{\GN^2}-J^2}
 = \frac{1}{\pi N_0N_6^2}\sqrt{{N_0^2N_6^2}-N_J^2} , \label{tl_y} \\
T^{\phi} &=
\frac{1}{2\pi J}\sqrt{\frac{P^2Q^2}{\GN^2}-J^2}
 =\frac{1}{2\pi  N_J}\sqrt{{N_0^2N_6^2}-N_J^2}. \label{tl_phi}
\end{align}

\section{Microscopic Entropy}
\label{entropy}

Using (\ref{cardy}),
either from (\ref{ry_central_charge}) and (\ref{tl_y})
or from (\ref{rphi_central_charge}) and (\ref{tl_phi}),
we finally obtain the microscopic entropy as 
\begin{align}
S_{\mathit{micro}}
&= \frac{\pi^2}{3}c^{y}T^{y} = \frac{\pi^2}{3}c^{\phi}T^{\phi} \nonumber\\
&= \frac{2\pi}{\hbar}\sqrt{\frac{P^2Q^2}{\GN^2}-J^2}
 = \pi\sqrt{{N_0^2N_6^2}-N_J^2},
\label{entropy_micro}
\end{align}
which exactly agrees with each other and
with the one derived macroscopically, (\ref{Sbh_extremal}).
Note that the Cardy formula will be valid for $T \gg 1$,
which is satisfied when $N_0 N_6 \gg N_J$ for $T^\phi$.
For other cases, like the Kerr/CFT correspondence \cite{Guica:2008mu},
the Cardy formula is not guaranteed to be applicable although
we hope it is.

Usually the Cardy formula is written as $S=2\pi\sqrt{\frac{cL_0}{6}}$. 
Therefore it is valuable to calculate the corresponding 
level of the Virasoro $L_0$. 
For (\ref{ry_central_charge}) and (\ref{tl_y}), this can be written as 
\begin{align}
L_0^{y} = \frac{\pi^2}{6}c^{y}(T^{y})^2
= \frac{R(P^2 Q^2-\GN^2J^2)}{4\GN\hbar P^2Q} 
= \frac{N_0^2N_6^2-N_J^2}{2N_0 N_6^2}. 
\end{align}
Similarly for (\ref{rphi_central_charge}) and (\ref{tl_phi}), we obtain 
\begin{align}
L_0^{\phi} = \frac{\pi^2}{6}c^{\phi}(T^{\phi})^2
= \frac{P^2Q^2-G_4^2J^2}{2\GN^2\hbar J}
= \frac{N_0^2N_6^2-N_J^2}{4N_J}.
\end{align}

\section{Conclusion and Discussion}
\label{coclusion_and_discussion}

In this paper, we calculated the entropy of the extremal 
rotating Kaluza-Klein black holes microscopically
by using Brown-Henneaux's method. 
Following \cite{Guica:2008mu}, we imposed
appropriate boundary conditions on the near horizon geometry 
of the black holes and then identified the diffeomorphisms
which preserve the boundary conditions
with the generators of the Virasoro algebras. 
Then by calculating the Dirac brackets of the corresponding
conserved charges, 
we determined the Virasoro algebras with non-vanishing central charges.
At the same time, we relate the physical parameters of the black holes
with the quantum numbers on the near horizon geometry.
Then we determined the temperatures of the dual chiral $CFT_2$'s.
From the central charges and the temperatures, 
by using the Cardy formula, we calculated the entropy of the extremal rotating 
Kaluza-Klein black holes microscopically,
which agrees with the one obtained macroscopically.

The rotating Kaluza-Klein black holes are known to be related 
to spinning D0-D6 bound states and we can also calculate the 
entropy from the D-brane viewpoint. Therefore we expect 
that we can obtain a deeper understanding of the chiral $CFT_2$'s
by considering a relation between our calculation and that by using 
the D-brane configuration.

\begin{center}
\noindent {\bf Acknowledgments}
\end{center}
We would like to thank Andrew Strominger for valuable comments.
T.~A. is supported by the Japan Society for the Promotion of Science (JSPS).
S.~T. is partly supported by
the Japan Ministry of Education, Culture, Sports, Science and
Technology (MEXT).
This work was supported by the Grant-in-Aid for
the Global COE Program ``The Next Generation of Physics, Spun from 
Universality and Emergence'' from the MEXT.

\vskip2mm

\begin{center}
\noindent {\bf Note added}
\end{center}
As this article was being completed, 
we received the preprint \cite{Lu:2008jk}.
In that paper,
the Kerr/CFT correspondence is applied to the higher dimensional Myers-Perry 
black holes and the Kerr-AdS black holes.


\providecommand{\href}[2]{#2}\begingroup\raggedright\endgroup

\end{document}